# Extraordinary Optical Transmission Induced by Excitation of a Magnetic Plasmon Propagation Mode in a Diatomic Chain of Slit-hole Resonators

H. Liu[1,]*, T. Li[1], Q. J. Wang[1], Z. H. Zhu[1], S. M. Wang[1], J. Q. Li[1], S. N. Zhu[1], Y. Y. Zhu[1] and X. Zhang[2]

[1]Department of Physics, National Laboratory of Solid State Microstructures, Nanjing University, Nanjing 210093, People's Republic of China

[2]5130 Etcheverry Hall, Nanoscale Science and Engineering Center, University of California, Berkeley, California 94720-1740, USA

Abstract

In this paper, we will propose that magnetic resonance nanostructures in a metal surface could be used to realize extraordinary optical transmission (EOT). Toward this goal, we designed and fabricated a one-dimensional diatomic chain of slit-hole resonator (SHR). Due to the strong exchange current interaction, a type of magnetic plasmon (MP) propagation mode with a broad frequency bandwidth was established in this system. Apparent EOT peaks induced by the MP mode were observed in our measured spectra at infrared frequencies. The strongest EOT peak was obtained at 1.07eV with an incident angle of $20^0$. The measured dependence of EOT peaks on the incident angle coincided with the theoretical results quite well. This proposed MP propagation mode in SHR structure has good potential applications in multi-frequency nonlinear optical processes.

*Author to whom correspondence should be addressed: liuhui@nju.edu.cn
Website of author's group: http://dsl.nju.edu.cn/dslweb/images/plasmonics-MPP.htm

Surface plasmon polariton (SPP) results from the coupling of electromagnetic wave and the collective electronic excitations on metal surface. It has considerably attracted the interest of researchers because of its important role in the extraordinary optical transmission (EOT) phenomena as reported by Ebbesen in 1998 [1-3]. According to recent studies, although SPP excitation is not the only mechanism responsible for EOT [4], it has undoubtedly provided one of the most remarkable contributions in the interaction between light and metallic nanostructures. Moreover, it has very important applications on multi-frequency nonlinear optical processes, such as biosensors [5], nanolasers [6], and *spaser* [7-8].

On the other hand, in 1999, Pendry reported that a nonmagnetic metallic element, such as a split ring resonator (SRR), with size below the diffraction limit, exhibits strong magnetic response and behaves like an effective negative permeability material [9]. Analogous to the surface plasmon resonances in metal nanoparticles, an effective media made of SRRs could support magnetic plasmon (MP) resonance [10-11] and lead to negative refraction [12]. According to the Babinet principle, the complementary structure of SRR (CSSRR) can also act as a good resonator in microwave range [13]. Besides SRR and CSRR, some new structures, such as nanorods [14], fish-nets [15], nano-sandwiches were invented to realize magnetic resonance at optical frequencies. Aside from the negative refraction, the application of MP has also been extended to cloaking materials [16] and polarization switches [17-18]. However, compared with SPP, which has a broad frequency bandwidth, MP is excited in a very narrow frequency range around the resonance frequency. This badly restricts MPs' applications in nonlinear optical processes which usually require multi-frequency excitations in a broad bandwidth.

In our recent theoretical study [19], a new kind of MP propagation mode in a linear monatomic chain of SRRs has been proposed to transfer EM wave signal in a subwavelength waveguide. In this paper, we will experimentally explore whether or not this MP propagation mode could lead to EOT. To realize this objective, we designed and fabricated a one-dimensional diatomic chain of slit-hole resonator (SHR). The MP propagation mode is expected to be established in this system with the help of the strong interaction between SHRs. Contrary to traditional uncoupled SRRs or CSRRs, the excitation of this MP propagation mode's optical branch could be tuned continually in a broad frequency bandwidth by adjusting the incident angles. The related EOT peak was observed in the transmitted spectrum at infrared frequencies. Results show that the measured dependence of the EOT peak on the incident angle coincides with the theoretical calculations quite well. This type of broadband MP mode has good potential applications in many multi-frequency nonlinear optical processes, such as fluorescence and Raman effects.

Figure 1(a) presents our designed slit-hole resonator structure, which is based on the designing idea proposed by the reference [13]. It comprises of two parts: a nano-hole near the edge of a semi-infinite golden film and a slit linking the hole with the edge. Its geometry parameters are also provided in this figure. Compared with SRR, SHR is much easier to be fabricated and its resonance frequency can reach infrared range. To study the EM response of the proposed SHR, we performed a set of finite-difference time-domain (FDTD) calculations using a commercial software package, the CST Microwave Studio (Computer Simulation Technology GmbH, Darmstadt, Germany). In addition, we relied on the Drude model to characterize the bulk metal properties. The metal permittivity in the infrared spectral range is given by $\varepsilon(\omega) = 1 - \omega_p^2 / (\omega^2 + i\omega\omega_\tau)$, where $\omega_p$ is the bulk plasma frequency, and $\omega_\tau$ is the relaxation rate. For gold, the characteristic frequencies fitted for experimental data are $\hbar\omega_p = 9.02 eV$ and $\hbar\omega_\tau = 0.027 eV$ [20]. In the numerical calculations, a near-field dipole source excited the SHR, and a probe was used to detect the local magnetic field at the center of the nano-hole. In the simulations, we observe a well-pronounced resonance frequency at $0.43 eV$. The oscillation of both the electric and magnetic fields at this resonance frequency is given in the online animation files [21]. The results show that the electric field is confined within the slit, while the magnetic field is concentrated inside the nano-hole. The SHR can be seen as an equivalent LC circuit, with the nano-hole working as a conductor, and the slit working as a capacitor (Fig.1 (b)). The induced resonance current in this LC circuit is also obtained in our simulations [21]. Due to the skin effect in the metal material, the current is only at the thin layer (thickness ~30nm) around the nano-hole.

The whole structure (SHR) could be seen as a magnetic dipole when the oscillation current is induced by an external wave at resonance frequency. We used a semi-analytic theory based on the Lagrangian formalism to describe the oscillation of this magnetic dipole. If $Q$ is the total oscillation charge in the SHR, L is the inductor of the nano-hole, and C is the capacitance of the slit, then we can write the Largangian equation of the system as $\Im = \frac{L\dot{Q}^2}{2} - \frac{Q^2}{2C}$. From the Euler-Lagrangian equation $\frac{d}{dt}\left(\frac{\partial \Im}{\partial \dot{Q}}\right) - \frac{\partial \Im}{\partial Q} = 0$, we can then obtain the oscillation equation of SHR as $\ddot{Q} + \frac{1}{LC}Q = 0$. If we define the SHR as a single magnetic dipole, $\mu = \dot{Q}/S$, where S is the circle area of the SHR, then $\ddot{\mu} + \omega_0^2 \mu = 0$, where $\omega_0^2 = 1/(LC)$ ($\omega_0$ is the resonance frequency of the SHR.

Based on the above SHR, a one-dimensional chain of magnetic resonator could be formed by connecting such a structure one by one. In our former work, a monatomic chain of SRRs is proposed and the MP mode is found in such a system [19]. However, its dispersion relation curve lies below the light line. At a given photon energy, the wave vector is not conserved when the photon is transformed into the MP mode. Such MP mode could not be excited by a far-field incident wave, and the EM energy could not be radiated out from the chain either. We conclude, therefore, that contrary to what we expected, the MP mode in a monatomic chain could not lead to EOT. In order to satisfy the wave vector matching condition, a diatomic chain of SHR is designed and presented in Fig. 2(a). Here, we can see that the unit cell of this chain is composed of two SHRs with different geometry sizes. The bigger SHR has the same structure as that given in Fig. 1(a), where the resonance frequency is obtained at $\omega_1 = 0.43 eV$. For the smaller SHR, the same simulation method is used and its resonance frequency is determined at $\omega_2 = 0.65 eV$.

For the infinite diatomic chain of SHRs, the equivalent LC circuit is given in Fig. 2(b). The Lagrangian equation of this system is expressed as:

$$\mathfrak{J} = \sum_m \left( \frac{L_1 \dot{Q}_m^2}{2} + \frac{L_2 \dot{q}_m^2}{2} - \frac{(Q_m - q_{m-1})^2}{2C} - \frac{(Q_m - q_m)^2}{2C} \right). \quad (1)$$

Here, we define the oscillating charges in the m-th unit cell as $Q_m$ for the bigger SHR with an inductor $L_1$, and $q_m$ for the smaller SHR with an inductor $L_2$ (m=0, ±1, ±2, ±3, …). The two corresponding magnetic dipoles, $U_m$ and $\mu_m$, are defined as $U_m = \dot{Q}_m / S$, $\mu_m = \dot{q}_m / s$, where S and s are the areas of bigger and smaller SHRs, respectively. From the Euler-Lagrangian equations $\frac{d}{dt}\left(\frac{\partial \mathfrak{J}}{\partial \dot{U}_m}\right) - \frac{\partial \mathfrak{J}}{\partial U_m} = 0$ and $\frac{d}{dt}\left(\frac{\partial \mathfrak{J}}{\partial \dot{\mu}_m}\right) - \frac{\partial \mathfrak{J}}{\partial \mu_m} = 0$ (m=0, ±1, ±2, ±3, …), we obtain the oscillation equations of the m-th bigger SHR and smaller SHR as:

$$\begin{cases} \ddot{U}_m + \omega_1^2 \cdot (2U_m - \mu_m - \mu_{m-1}) = 0 \\ \ddot{\mu}_m + \omega_2^2 \cdot (2\mu_m - U_m - U_{m+1}) = 0 \end{cases}, \quad (2)$$

where $\omega_1 = 1/\sqrt{L_1 C}$, and $\omega_2 = 1/\sqrt{L_2 C}$. We seek a general solution of Equation (2) in the form of the magnetic plasmon (MP) wave,

$$\begin{cases} U_m = U_0 \cdot \exp(i(\omega t - k \cdot md)) \\ \mu_m = \mu_0 \cdot \exp(i(\omega t - k \cdot (md + d/2))) \end{cases}, \quad (3)$$

where $\omega$ and $k$ are the angular frequency and wave vector, respectively, $U_0$ and $\mu_0$ are the initial values of the magnetic dipole moment at $m = 0$, and $d = 650 nm$ is the period of the chain. By substituting Equation (3) into Equation (2), and then solving the Eigen equations for $U_0$ and $\mu_0$, the MP dispersions are attained as:

$$\omega_\pm^2 = (\omega_1^2 + \omega_2^2) \pm \sqrt{(\omega_1^4 + \omega_2^4) + 2\omega_1^2 \omega_2^2 \cos(kd)}. \quad (4)$$

The dispersion relations are numerically depicted as two solid black curves in Fig. 3 (a). There are two separate dispersion branches for the diatomic chain: upper branch $\omega_+(k)$ and lower branch $\omega_-(k)$. For these two branches, the respective resonant manners of the m-th unit cell are quite different. Our simulations [21] show that for the lower branch, $\omega_-(k)$, $U_m$ and $\mu_m$ oscillate in the same phase (see Fig.3 (b)), whereas for the upper branch, $\omega_+(k)$, they oscillate in the anti-phase (see Fig.3 (c)). Using the analogy of the diatomic model of crystal lattice wave [22], we can refer to the upper curve $\omega_+(k)$ as optical branch and the lower curve $\omega_-(k)$ as acoustic branch. Compared with the monatomic chain [19], which possesses only the acoustic dispersion branch, the optical branch is a kind of MP mode found in the diatomic chain. In Fig. 3 (a), the light line in the free space is given as a blue dotted straight line ($\omega = ck_0$). It is quite exciting to see that the upper optical branch intersects with the light line, and that the major part of this curve lies on the left side of the intersection point. For an oblique incident plane wave, the resonant excitation of the MP modes could be achieved under the wave vector matching condition:

$$k = k_0 \sin\theta, \qquad (5)$$

where $\theta$ is the incident angle as denoted in Fig. 2 (a). Combining Equations (4) and (5), the dependence of resonance excitation frequency on the incident angle could be solved numerically, and is shown as a white line in Fig. 4(b). For a perpendicular incident wave ($\theta = 0^0$), the MP mode is excited at the frequency $1.11eV$. At the crossing point of the optical branch curve and the blue line in Fig. 3 (a), the MP mode is excited by a plane wave propagating along the metal surface ($\theta = 90^0$), with the corresponding frequency at $0.924eV$. Thus, the excitation frequency range of the MP mode is $0.924 \sim 1.11eV$ with a bandwidth $0.186eV$

Based on the above theoretical model, we expected that the optical branch of the MP mode could lead to EOT when excited by the incident plane wave. To achieve this experimentally, a one-dimensional chain of SHRs on a silver film was fabricated with the focused-ion-beam (FIB) system (Strata FIB 201, FEI company, 30keV Ga ions). The sample picture is shown in Fig. 2(c), along with its structure parameters. The sample is set on a rotation table, and as can be seen, a polarized light is directed to the sample with its e-field in the y direction. The transmitted waves are collected by an optical-spectrum analyzer (ANDO AQ-6315A), and the measured transmission efficiency normalized to the area of the holes are shown in Fig. 4(a) under different incident angles. As expected, the EOT peaks are evidently observed around the resonance frequency of the optical MP mode, when the incident angle is in the range of $6^0 \leq \theta \leq 40^0$. In contrast, there is no obvious peak to be found in the transmission curve for the small incident angles ($\theta < 6^0$). This is because MPs EOT is formed via the coupling to the h-field component of light. In order to excited strong magnetic resonance, magnetic flux must pass through the nano-hole, and the normal component of the h-field must not be equal to zero ($H_z \neq 0$). When the incident angle $\theta$ is less than $6^0$, most of the h-field of the incident wave would be parallel to the metal film and, therefore, would not be able to pass through the nano-hole. There is not enough energy coupling into the MP mode. Hence, it is hard to find obvious EOT peaks in our measurement. In order to induce more magnetic flux to pass through the nanohole, we increased the incident angle. Consequently, more EM energy was absorbed into the MP modes and then radiated out from the backside of the sample. The corresponding EOT peaks became stronger and were then measured in the transmitted wave (Fig. 4(a)). Afterwards, we attained the strongest EOT peak at around $\theta = 20^0$. When the incident angle was further increased, the EOT peak declined and became too weak to be observed when $\theta > 40^0$. This is because if the incident angle $\theta$ is increased, the projected area ($S'$) of the fabricated region ($S$) in the incident direction will be decreased ($S' = S\cos\theta$). Less EM energy could be coupled into SHRs and attributed to the EOT peak. Thus, no evident EOT peak could be measured in the range $\theta > 40^0$. For the incident angle $8^0 < \theta < 32^0$, the normalized transmission efficiency could exceed unity (see Fig.4 (a)), proving that the excitation of the optical MP mode leads to extraordinary transmission. The angle dependence of the EOT peak obtained in this experiment is quite different from that of the SPP mode.

The excitation of SPP is through the coupling to the e-field component of light and the EOT peak could be attained with the perpendicular incident wave ($\theta = 0$).

In order to obtain the comparison between these measured experiment results and the theoretical results, the transmission curves under different incident angles were combined into a 2-D contour map (Fig. 4(b)). In this map, the transmitted intensity is denoted as the brightness of each point. From the figure, it could be seen that the bright part of the map matches the theoretical white line quite well. This confirms that the measured EOTs are obtained from the excitation of the optical MP modes in the diatomic chain of SHRs. Actually, the bandwidth of the optical branch can be enlarged if we increase the coupling interaction between elements by changing the length of the slit. In our experiments, we fabricated another sample with the slit length equals 50nm (smaller than 70nm for the old sample). Other structure parameters are kept the same value as Fig. 2 (a). The bandwidth obtained is about 0.21eV, which is larger than the bandwidth of the old sample. The above EOT based on 1-D diatomic SHR chain could be also realized in 2-D SHR structures [21]. The detailed discussion will be provided in another paper.

From the above experimental results, we can see that the MP propagation mode in our system can be excited in a broad frequency bandwidth. Actually, the dispersion curves in Fig. 3 are divided into two parts by the blue light line. The part above the blue line is the bright MP mode, which can couple to the far-zone optical field. Aside from the EOT reported in this paper, the bright MP mode could also be used to produce efficient nanolasers, which has recently aroused intense interest [6]. Meanwhile, the rest part below the blue line corresponds to the dark MP mode, which cannot be excited by the far-field wave and whose energy does not radiate outwards. Without radiation loss, the dark MP mode can be greatly amplified by the stimulated emission from an active medium (e.g., quantum dots and the like) similarly as SPASER achieved in dark SPP mode [7-8]. This could provide a good nanoscale optical source for numerous potential applications in nonlinear optical processes, such as single-molecule detection and florescence imaging.

In summary, we proposed and studied a magnetic plasmon propagation mode with a broad frequency bandwidth. The optical branch of this MP mode was excited by an oblique incident wave, the occurrence of which led to an extraordinary transmission of infrared light in our experiments. The measured dependence of EOT peaks on the incident angle coincided with the theoretical results quite well. Both the bright and dark modes could be obtained in this system. Their numerous prospective applications in nonlinear optical processes are worthy to be explored in the future.

This work is supported by the National Natural Science Foundation of China (No.10604029, No.10704036 and No.10874081), and by the National Key Projects for Basic Researches of China (No.2009CB930501, No.2006CB921804 and No. 2004CB619003).

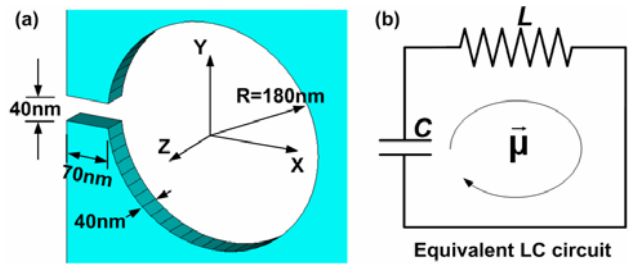

Fig.1 (a) Structure of a single SHR; (b) Equivalent LC circuit of the single SHR.

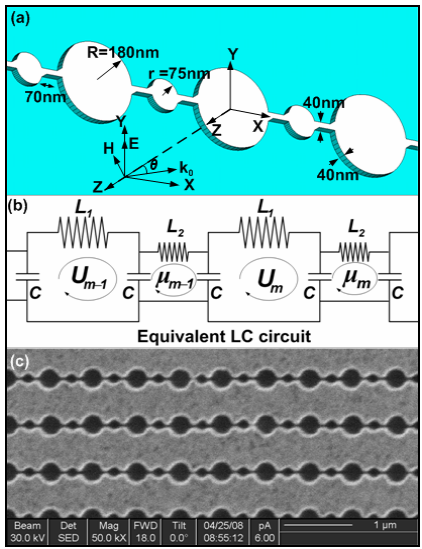

Fig.2 (a) Structure of a diatomic chain of SHRs; (b) Equivalent LC circuit of the chain; (c) FIB image of the fabricated sample.

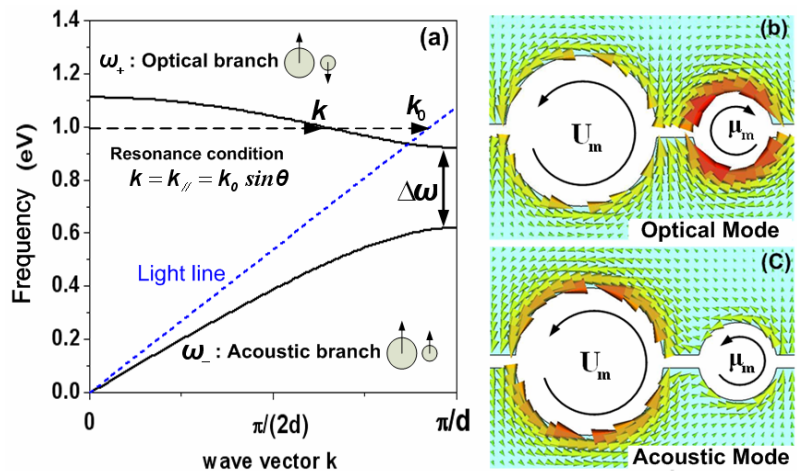

Fig.3 (a) Dispersion curves for the MP modes in the diatomic chain of SHRs; the local current distributions are calculated for the (b) optical and (c) acoustic modes.

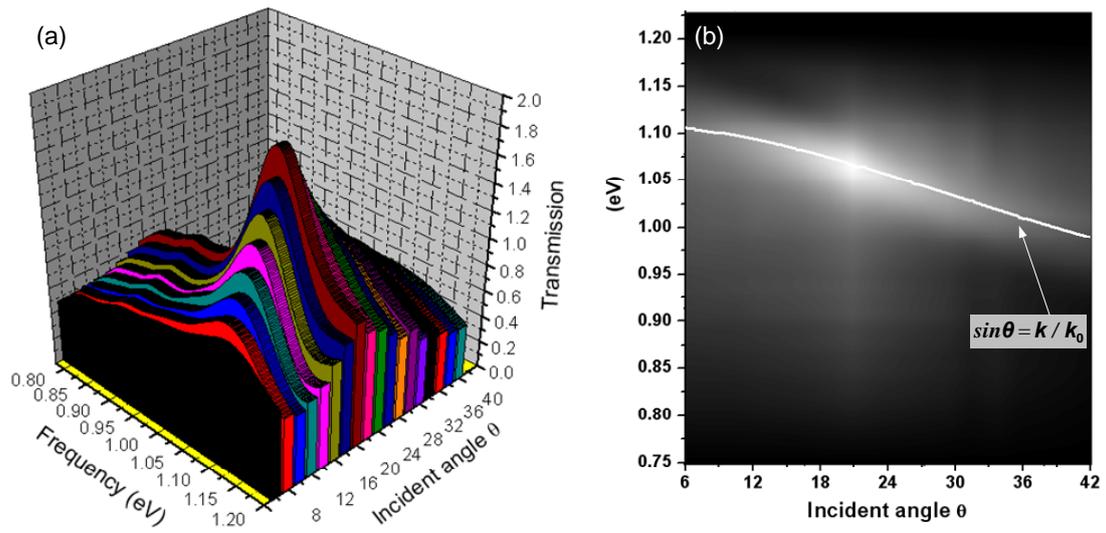

Fig.4 (a) Measured transmission efficiency normalized to the area of the holes under different incident angles; (b) Measured transmission map and the calculated angular dependence curve of the optical MP mode.